\documentclass[conference]{IEEEtran}
\IEEEoverridecommandlockouts
\usepackage{cite}
\usepackage{amsmath,amssymb,amsfonts}
\usepackage{algorithmic}
\usepackage{graphicx}
\usepackage{textcomp}
\usepackage{xcolor}
\usepackage{soul}
\usepackage{multirow}
\usepackage{array}
\usepackage{fancyhdr}
\usepackage{hyperref}

\fancypagestyle{plain}{
  \fancyhf{}
  \fancyhead[C]{Conference on \LaTeX}     
  \fancyfoot[L]{This is a notice}

}
\usepackage{eso-pic}

\def\BibTeX{{\rm B\kern-.05em{\sc i\kern-.025em b}\kern-.08em
    T\kern-.1667em\lower.7ex\hbox{E}\kern-.125emX}}
\begin{document}
\bstctlcite{IEEEexample:BSTcontrol}
\title{A Performance Evaluation of Filtered Delay Multiply and Sum Beamforming for Ultrasound Localization Microscopy: Preliminary Results\\
}

\author{A. N. Madhavanunni$^{1}$, Niya Mariam Benoy$^{2}$, Mahesh Raveendranatha Panicker$^{3}$, Himanshu Shekhar$^{4}$\\
\textit{$^{1}$Indian Institute of Technology Palakkad, Kerala, India,}
     \textit{$^{2}$College of Engineering Trivandrum, Kerala, India,}\\
      \textit{ $^{3}$Singapore Institute of Technology, Singapore, }
    \textit{ $^{4}$Indian Institute of Technology Gandhinagar, Gujarat, India}\\
    \small $^{1}$121813001@smail.iitpkd.ac.in, $^{2}$niyamariambenoy@gmail.com, $^{3}$mahesh.panicker@singaporetech.edu.sg, $^{4}$himanshu.shekhar@iitgn.ac.in}

\begin{titlepage}
    \vspace*{\fill}
\fontsize{15}{18}\selectfont\textcopyright { 2024 IEEE. Personal use of this material is permitted. Permission from IEEE must be obtained for all other uses, in any current or future media, including reprinting/republishing this material for advertising or promotional purposes, creating new collective works, for resale or redistribution to servers or lists, or reuse of any copyrighted component of this work in other works.}
    \vspace*{\fill}
\end{titlepage}

\AddToShipoutPictureBG{%
  \AtPageUpperLeft{%
    \setlength\unitlength{1in}%
    \hspace*{\dimexpr0.5\paperwidth\relax}
    \makebox(0,-0.75)[c]{{This work has been accepted in the IEEE South Asian Ultrasonics Symposium 2024 (IEEE SAUS 2024).}}%

}}



\maketitle

\begin{abstract}
Ultrafast ultrasound localization microscopy (ULM), which has shown promising results in microvascular imaging, overcomes the typical trade-off between resolution and penetration depth. Combining ultrasound contrast agents and high frame rate imaging enables ULM to visualize microvasculature and quantify flow. However, the quality of the microvascular maps obtained depends on the signal-to-noise ratio of the received signals, image reconstruction techniques, and the microbubble (MB) localization and tracking algorithms used. Most reported research in ULM employs the conventional delay and sum (DAS) beamforming technique for image reconstruction despite its limited contrast and resolution. In this work, a filtered delay multiply and sum (F-DMAS) beamforming approach with non-steered plane wave transmit was employed for ULM, and its performance was compared with the conventional DAS-based approach for the different localization algorithms available in the Localization and Tracking Toolbox for Ultrasound Localization Microscopy. We also introduce two novel image quality measures that can overcome the limitations of conventional quality metrics that require suitable targets for evaluation. We also report the preliminary \textit{in-vitro} investigation of F-DMAS with B-mode and power Doppler maps for microvascular imaging. The results are promising with enhanced contrast and lateral resolution, and suggest that further experimental studies are warranted.
\end{abstract}
\section{Introduction}
\label{sec:intro}
Ultrasound imaging is extensively used as a diagnostic imaging technique because of its ease of use, affordability, and non-ionizing nature when compared to other imaging modalities. It is an effective non-invasive tool for soft-tissue examinations, guided interventions, and investigation of blood flow dynamics at high frame rates \cite{Tanter2014UltrafastUltrasound}. Contrast-enhanced ultrasound (CEUS) imaging enhances the contrast from blood-filled organs by employing intravenously injecting microbubbles into the bloodstream \cite{frinking2020three}. However, the resolution of conventional ultrasound imaging is limited by the diffraction limit, making microvascular imaging challenging.

In the past decade, the combined use of ultrasound contrast agents and high frame rate imaging, ultrasound localization microscopy (ULM), has enabled sub-wavelength imaging and microvasculature visualization. It overcomes the resolution-penetration trade-off and has the potential to revolutionize the diagnosis of stroke, arteriosclerosis, cancer, and other pathologies \cite{couture2018ultrasound, christensen2020super, demene2021transcranial}. The spatiotemporal sensitivity of ULM depends on accurate and robust microbubble localization, which relies on signal-to-noise ratios of the received signals, receive beamforming, microbubble localization algorithms, and rendering of their trajectories \cite{heiles2022performance}. 

Most of the reported research in the literature on ULM is focused on microbubble localization algorithms applied to the delay and sum (DAS) beamformed images. Also, there have been some efforts towards employing adaptive beamforming techniques for resolving ultrasound contrast microbubbles, which have shown better results than DAS beamforming \cite{corazza2022microbubble, tasbaz2021improvement, espindola2018adaptive, diamantis2018resolving}. However, adaptive beamformers are computationally complex and require tuning of multiple parameters for optimal results. Although novel beamformers have been investigated extensively for B-mode and flow imaging, studies on the influence of beamformers on the quality of ULM are limited. Moreover, evaluating contrast and resolution using conventional metrics requires suitable targets like cysts/lesions for contrast and point targets for resolution measurement. However, in the context of ULM, such targets would be absent, and hence conventional metrics won't be robust. In this regard, the contributions of this work are two-fold. 1) Improving the resolution and contrast of ULM by employing filtered delay multiply and sum (F-DMAS) beamforming, a technique that doesn't require parameter tuning and has a lower computational overhead than most adaptive beamformers. To our knowledge, this is the first attempt to investigate the F-DMAS beamformer for ULM. 2) We also introduce two novel metrics to quantify the contrast and resolution in terms of local contrast and lateral spread for evaluating the performance of the beamformers with different localization techniques.

%



%
\begin{figure*}[!t]
\centering
{\includegraphics[width=1\textwidth]{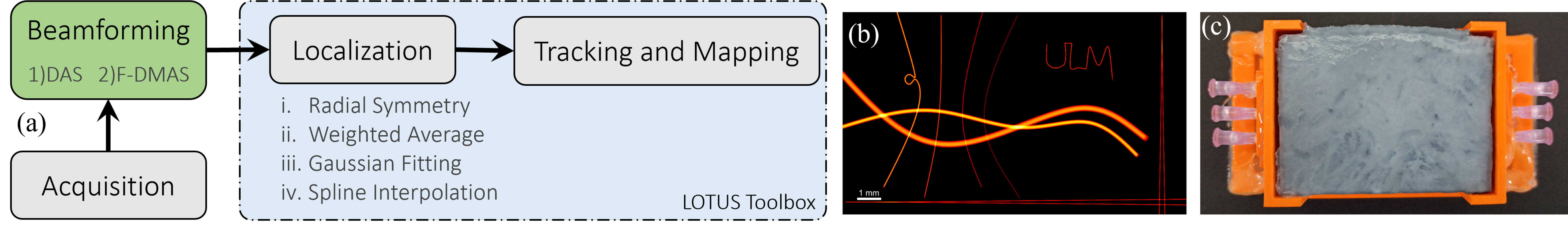}}
\caption{(a) ULM pipeline employed for the study (b) Ground truth image of the simulated phantom (c) \textit{In-vitro} microvascular PVA phantom} 
\label{fig_ULMpipeline}
\end{figure*}

\section{Materials and Methods} \label{sec:methods}


This section describes the simulated and experimental datasets, the steps involved in the ULM pipeline as shown in Fig. \ref{fig_ULMpipeline}(a), and the proposed evaluation metrics. 
\subsection{\textit{In-silico} and \textit{In-vitro} Datasets}
A microvascular flow phantom that had 11 tubes with different geometries and complexity (Fig. \ref{fig_ULMpipeline}(b)) was simulated in Field II \cite{Jensen1992CalculationTransducers, jensen1996field}. The phantom media was adopted from the \textit{in-silico} angiography data in the PALA open dataset \cite{heiles2022performance}. A 128-element linear array of $0.11\ mm$ pitch was used for insonification with non-steered plane waves at a center frequency of $15.625\ MHz$ and a frame rate of $500\ Hz$. A speed of sound of $1540\ m/s$ was used for simulation and the received echoes were sampled at $100\ MHz$. 

For the experimental investigations, a poly-vinyl-alcohol (PVA) based microvascular flow phantom having vessels of $110\ \mu m$ diameter was prepared using a previously reported protocol \cite{mercado2018effect} (Fig. \ref{fig_ULMpipeline}(c)). The raw data was acquired with a Verasonics Vantage 128 research ultrasound system using a 128-element linear array (L11-5v) of center frequency $7.6\ MHz$ and $1.5\lambda$ pitch. Non-steered plane waves at a frame rate of $2000\ Hz$ were used for insonification and the received echoes were sampled at $31.25\ MHz$. A 3\% (weight/volume) potato starch solution with a liquid base of water-glycerol mixture at a volume/volume ratio of 53\%-44\% was used as the blood-mimicking fluid to enhance contrast in microvascular imaging \cite{arunKumar2024Echogenicity}. A singular value decomposition (SVD) was used to remove the clutter components.

\subsection{Beamforming}
Without loss of generality, non-steered plane wave transmission with an $N_c$-element linear ultrasonic array was considered and $u_i[n]$ represents the received echo signal at the $i^{th}$ element unless otherwise specified. 

\subsubsection{Delay and Sum (DAS) beamforming}

For a pixel, $p$ at $\left[x_p,\ z_p\right]$, the DAS beamformed signal \cite{Perrot2021SoBeamforming} can be mathematically expressed as:
\begin{equation}\label{dasEq}
    y_{DAS}\ [p]=\ \sum_{i\ =\ 1}^{N_c}{w_i\left[p\right] u_i[\Delta n_p]}
\end{equation}
where, $\Delta n_p = \sqrt{(x_i-x_p)^2 + (z_p)^2}$ is the total ultrasound propagation delay for the pixel, $p$ and $w_i\left[p\right]$ is the apodization weight.

\subsubsection{Filtered Delay Multiply and Sum (F-DMAS) beamforming}\label{fdmas}

The delay multiply and sum (DMAS) algorithm involves the multiplication of delay-compensated signals from every channel with that of every other channel as in (\ref{dmasEq}). The DMAS beamformed signal for the pixel, $p$ at $[x_p,\ z_p]$ is given by:
\begin{equation}\label{dmasEq}
    \hat{y}_{DMAS}\ \left[p\right]=\ \sum_{i=1}^{N_c-1}\sum_{j=i+1}^{N_c}{v_i\left[n-\Delta n_p\right]\ \times\ v_j\left[n-\Delta n_p\right]}
\end{equation}
where, $v_i[n] = sign(u_i[n]) \times \sqrt{|u_i[n]|}$ \cite{Matrone2015TheImaging}. The frequency spectrum of $y_{DAS}$ in (\ref{dasEq}) is a band centered around the transmit center frequency ($f_c$) whereas for ${\hat{y}}_{DMAS}$ in (\ref{dmasEq}), the spectrum has two major components, one centered at baseband and the other at ${2f}_c$ because of the multiplication of signals with similar frequency components \cite{Matrone2015TheImaging}. In this regard, a bandpass filter (BPF) with the passband centered around ${2f}_c$ was used to eliminate the zero-frequency components and to obtain the F-DMAS beamformed signal, $y_{DMAS}$. 


%
\begin{figure*}[!t]
\centering
{\includegraphics[width=0.8\textwidth]{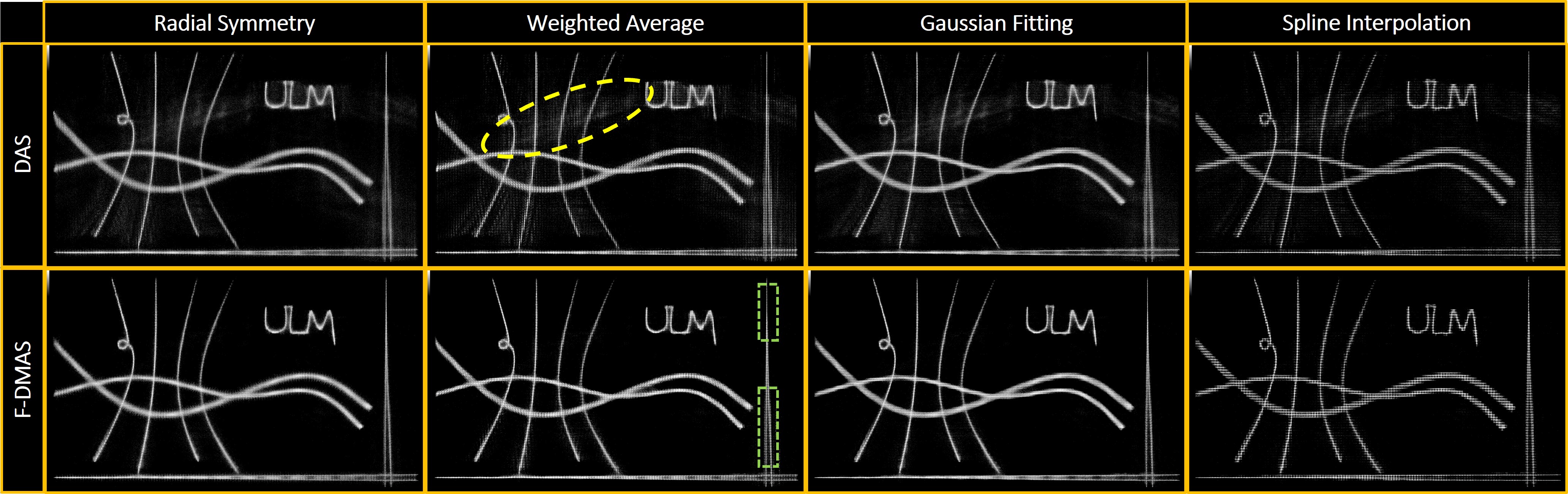}}
\caption{Localization maps obtained using different localization algorithms with DAS and F-DMAS beamforming for \textit{in-silico} datasets}
\label{fig_locMap}
\end{figure*}
\begin{figure*}[!t]
\centering
{\includegraphics[width=0.8\textwidth]{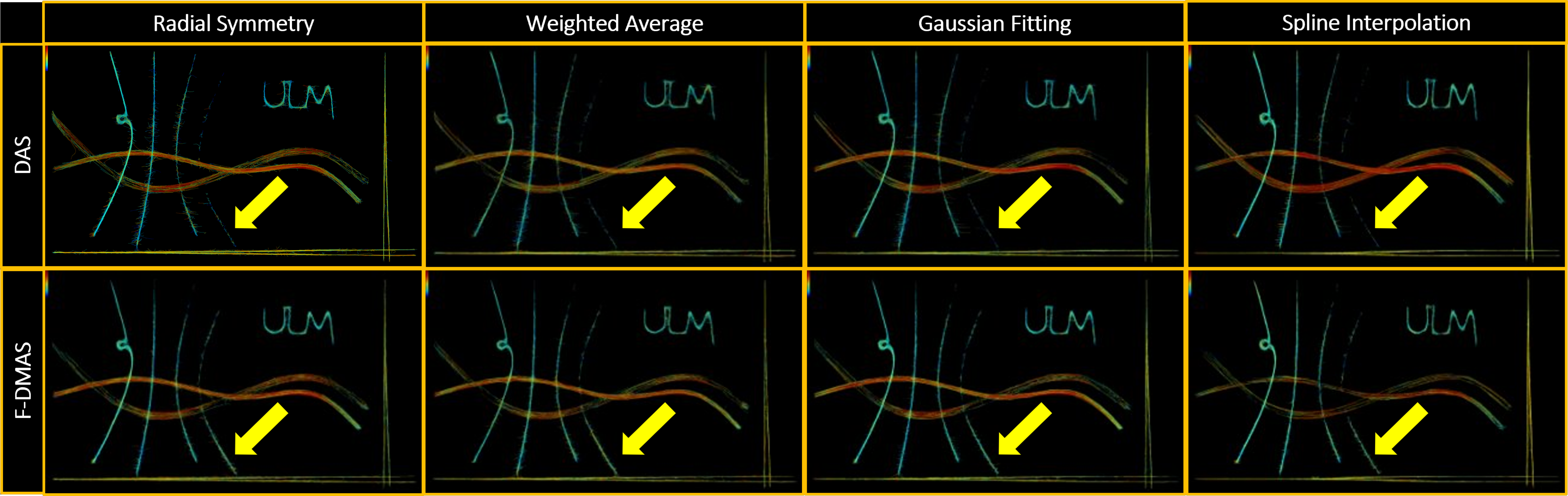}}
\caption{Normalized velocity maps obtained with DAS and F-DMAS beamforming for \textit{in-silico} datasets}
\label{fig_normVelMap}
\end{figure*}

\subsection{Localization, Tracking, and Mapping}
The localization and tracking of MBs were performed using the Localization and Tracking Toolbox for Ultrasound Localization Microscopy (LOTUS) \cite{heiles2022performance}. In this study, we employed four different localization algorithms based on Spline Interpolation (Sp-Interp), Gaussian Fitting (Gauss-Fit), Radial Symmetry (RS), and Weighted Average (WA) that have been reported in \cite{heiles2022performance}. The localization was performed on the cropped beamformed images (having a pixel resolution of $\lambda\times\lambda$ that were centered on the highest-intensity pixel.


\textit{1) Interpolation:} The image was interpolated to a resolution of $0.1\lambda \times 0.1\lambda$ using the spline method \cite{heiles2022performance, song2018effects}.

\textit{2) Gaussian Fitting:} A Gaussian kernel was used for fitting the MB and the peak intensity point is estimated \cite{heiles2022performance, song2018effects}.
  
\textit{3) Weighted Average:} The localization of the MB centroid was estimated as a weighted average of the intensity in a horizontal and a vertical line. For instance, the lateral position was estimated by performing a weighted summation of the image intensity (without log compression) over the 5 axial pixels using a weight vector of  [-2; -1; 0; 1; 2] and dividing it by the total intensity \cite{heiles2022performance}.

\textit{4) Radial Symmetry:} The radial symmetry-based localization algorithm, introduced in optical super-resolution microscopy \cite{parthasarathy2012rapid}, exploits the shape of the point spread function (PSF) and estimates the best-fit radial symmetry center. It is one of the most reliable techniques to localize MBs \cite{hingot2019development}.

Once the sub-wavelength positions were obtained with the above four localization techniques, tracking, velocimetry, and mapping were performed using the LOTUS toolbox and its associated MATLAB scripts \cite{heiles2022performance}.

\subsection{Evaluation Metrics}
The conventional contrast and resolution measures like contrast-to-noise ratio and full-width half maximum (FWHM) are typically evaluated on cyst targets and point targets, respectively. However, in the context of ULM wherein the microvasculature is of prime interest, suitable targets for the conventional contrast and resolution measures would be absent in the image. We introduce two novel metrics to quantify the contrast and resolution in terms of local contrast and lateral spread for evaluating the performance of the beamformers with different localization techniques.

\textit{1) Local contrast score:} Motivated by the idea of multi-scale transform and contrast images applied to X-ray images in \cite{schaetzing2007agfa}, we introduce a metric to measure the local contrast feature for ULM. Towards this, a local standard deviation image \cite{Sezn1989tmi} is estimated by computing the root-mean-square (RMS) value using a moving kernel over the localization map. Since the sub-wavelength features are of prime interest in ULM, the kernel size was chosen as $2\times2$ which corresponds to $0.2\lambda \times 0.2\lambda$, where $\lambda$ is the wavelength of the transmitted ultrasound signal. The local contrast score is reported as the mean and standard deviation (Std. Dev.) of the local standard deviation image.

\textit{2) Lateral spread score:} To determine the spread of the canals, we employed an extended version of a typical FWHM function. It is calculated as the mean width of the main lobe at its half maximum from the normalized intensity plot for the selected channels. For this study, to estimate the lateral spread, the vertical canal as shown in green in Fig. \ref{fig_locMap} was selected, and the mean FWHM along the canal was estimated as a function of $\lambda$.


\section{Results and Discussion}
\label{sec:results}

The localization maps obtained using the selected four localization techniques with DAS and F-DMAS beamforming for the \textit{in-silico} dataset are shown in Fig. \ref{fig_locMap}. The localization maps for the F-DMAS demonstrated better lateral resolution than DAS (particularly for the vertical canals highlighted in dashed green rectangles in Fig. \ref{fig_locMap}), along with reduced side lobe artifacts (over the regions highlighted in dashed yellow in Fig. \ref{fig_locMap}). Moreover, the overall contrast of the microvasculature for F-DMAS was enhanced relative to DAS. The improvement in the contrast and lateral resolution can be explained by the following reasons: 1) The “synthetic” receive aperture widening due to the correlation function in (\ref{dmasEq}) and 2) the improved coherence after the cross multiplications in (\ref{dmasEq}).

\begin{table}[!t]
\centering
\caption{Comparison of local contrast score and the lateral spread score between the conventional DAS and F-DMAS based ULM} 
\vspace{4pt}
\label{tab:quant_metric}
\begin{tabular}{cccccc}
\hline
\hline
&& \multicolumn{2}{c}{Local Contrast Score}&  \\ 
\cline{3-4} 
\multirow{2}{*}{\textbf{\begin{tabular}[c]{@{}c@{}}Beamforming \\ Method \end{tabular}}} &
\multirow{2}{*}{\textbf{\begin{tabular}[c]{@{}c@{}}Localization\\ Method  \\  \end{tabular}}} &
 \multirow{2}{*}{\textbf{\begin{tabular}[c]{@{}c@{}}Mean \end{tabular}}} &
 \multirow{2}{*}{\textbf{\begin{tabular}[c]{@{}c@{}}Std. Dev.\end{tabular}}} &
 \multirow{2}{*}{\textbf{\begin{tabular}[c]{@{}c@{}}Lateral \\Spread Score\end{tabular}}} \\
   &    &    \\ \hline
\multirow{4}{*}{\textbf{\begin{tabular}[c]{@{}c@{}}DAS\end{tabular}}}& RS & 0.854 & 0.258 & $0.462 \lambda$ \\ 
& WA & 0.868 & 0.243 & $0.365 \lambda$ \\ 
& Gauss-Fit & 0.869 & 0.244 & $0.325 \lambda$ \\ 
& Sp-Interp & 0.905 & 0.213 & $0.263 \lambda$ \\ \hline
\multirow{4}{*}{\textbf{\begin{tabular}[c]{@{}c@{}}F-DMAS \end{tabular}}}& RS & 0.901 & 0.237 & $0.346 \lambda$ \\
& WA & 0.919 & 0.217 & $0.330 \lambda$\\ 
& Gauss-Fit & 0.919 & 0.219 & $0.319 \lambda$ \\ 
& Sp-Interp & 0.932 & 0.199 & $0.257 \lambda$ \\ \hline

Ground Truth & - & 0.952 & 0.173 & $0.238 \lambda$ \\ \hline
\hline
\end{tabular}
\end{table}


%
\begin{figure}
\centering
{\includegraphics[width=0.48\textwidth]{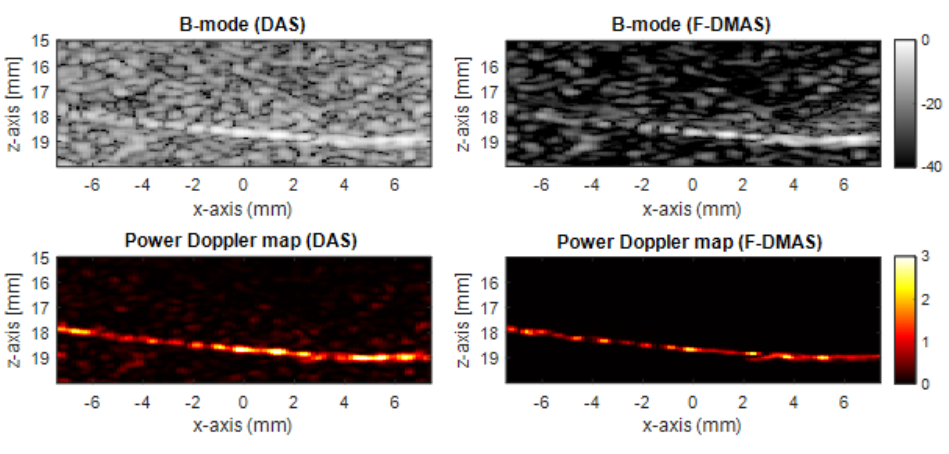}}
\caption{\textit{In-vitro} results: B-mode image and power Doppler map obtained for a microvascular phantom with DAS and F-DMAS beamforming}
\label{fig_invitroResults}
\end{figure}

The reported results were validated quantitatively with the contrast and resolution metrics and are presented in Table \ref{tab:quant_metric}. The quantitative metrics for DAS and F-DMAS methods were compared with those of the ground truth image. The lateral spread was calculated as the mean value over the vertical canal regions indicated in green in Fig. \ref{fig_locMap}. The local contrast score and the lateral spread obtained for the F-DMAS images were better than that of DAS irrespective of the localization technique. However, it was observed that the localization based on spline interpolation showed the best results (closest to the ground truth values) irrespective of the beamformer used. 

The normalized velocity maps obtained with the selected four localization techniques for DAS and F-DMAS beamforming are shown in Fig. \ref{fig_normVelMap}. The thinnest vertical channel (indicated with yellow marks in Fig. \ref{fig_normVelMap}) was completely missing with DAS beamforming while its velocity traces were obtained for F-DMAS images irrespective of the localization techniques. Overall, the qualitative results and quantitative metrics suggest that the F-DMAS beamforming improved the performance of all the localization techniques, even for the weighted average method, which is one of the most computationally efficient localization methods \cite{heiles2022performance}.

Furthermore, the efficacy of F-DMAS beamforming for ULM was evaluated \textit{in-vitro} and the corresponding B-mode images and power Doppler maps are shown in Fig. \ref{fig_invitroResults}. Better clutter suppression and improved resolution were observed with F-DMAS and are evident in the B-mode image and Power Doppler maps in Fig. \ref{fig_invitroResults}. These improvements can be attributed to “synthetic” receive aperture widening and the improved coherence in F-DMAS.


\section*{Acknowledgment}
\label{sec:thanks}

The authors would like to thank Dr. Olivier Couture, Sorbonne University, France, and Dr. Karla Mercado-Shekhar, IIT Gandhinagar, India, for the useful discussions. We acknowledge the PALA open dataset, the LOTUS toolbox, and the associated scripts for making the performance benchmarking easily possible. We also acknowledge the computing facilities provided by the  Centre for Computational Imaging, Indian Institute of Technology Palakkad, India.


\bibliographystyle{IEEEtran}
\small\bibliography{references.bib}

\end{document}